\RequirePackage{fix-cm}

\documentclass{svjour3}                     % onecolumn (standard format)
\smartqed  % flush right qed marks, e.g. at end of proof
\usepackage{amsmath,graphicx}
\usepackage{url}
\usepackage{multirow}
\usepackage{enumitem}
\usepackage{cite}
\usepackage{amssymb}
\usepackage{mathtools}
\usepackage{refcount}
\usepackage{booktabs}
\usepackage[numbers]{natbib}
\usepackage{subfigure} %FIGTOPCAP
\usepackage{epstopdf}
\DeclarePairedDelimiter\floor{\lfloor}{\rfloor}
\DeclarePairedDelimiter\abs{\lvert}{\rvert}%
\usepackage{xcolor}

\newcommand{\plh}{%
  {\ooalign{$\phantom{0}$\cr$\scriptstyle\times$\cr}}%
}

\usepackage{fancyhdr}
\begin{document}
\pagestyle{fancy}
\fancyfoot[C]{\footnotesize Preprint of Lin K.W.E., Balamurali B.T., Koh E., Lui S., Herremans D. In Press. Singing Voice Separation Using a Deep Convolutional Neural Network Trained by Ideal Binary Mask and Cross Entropy. \emph{Neural Computing and Applications}, Springer. }

\title{Singing Voice Separation Using a Deep Convolutional Neural Network Trained by Ideal Binary Mask and Cross Entropy\thanks{This work is supported by the MOE Academic fund AFD 05/15 SL and SUTD SRG ISTD 2017 129.}
}

\titlerunning{Singing Voice Separation Using a CNN Trained by IBM and Cross Entropy}        % if too long for running head

\author{Kin Wah Edward Lin$^\ddagger$ \and Balamurali B.T.\and Enyan Koh \and Simon Lui \and Dorien Herremans
}

%\authorrunning{Short form of author list} % if too long for running head

\institute{K.W.E Lin$^\ddagger$, Balamurali B.T., E. Koh, and S. Lui \at Singapore University of Technology and Design, Singapore \\
\email{edward\_lin@mymail.sutd.edu.sg$^\ddagger$, balamurali\_bt@sutd.edu.sg,\\enyan\_koh@mymail.sutd.edu.sg, simon\_lui@sutd.edu.sg}\\
$^\ddagger$ Corresponding Author 
\and
D. Herremans \at
Singapore University of Technology and Design, Singapore \\
\& Institute for High Performance Computing, A*STAR, Singapore \\
\email{dorien\_herremans@sutd.edu.sg}
}

\date{Received: 14/12/2018 / Accepted: 30/11/2018}
% The correct dates will be entered by the editor

\maketitle

\begin{abstract}

Separating a singing voice from its music accompaniment remains an important challenge in the field of music information retrieval. We present a unique neural network approach inspired by a technique that has revolutionized the field of vision: pixel-wise image classification, which we combine with cross entropy loss and pretraining of the CNN as an autoencoder on singing voice spectrograms. The pixel-wise classification technique directly estimates the sound source label for each time-frequency (T-F) bin in our spectrogram image, thus eliminating common pre- and postprocessing tasks. The proposed network is trained by using the Ideal Binary Mask (IBM) as the target output label. The IBM identifies the dominant sound source in each T-F bin of the magnitude spectrogram of a mixture signal, by considering each T-F bin as a pixel with a multi-label (for each sound source). Cross entropy is used as the training objective, so as to minimize the average probability error between the target and predicted label for each pixel. By treating the singing voice separation problem as a pixel-wise classification task, we additionally eliminate one of the commonly used, yet not easy to comprehend, postprocessing steps: the Wiener filter postprocessing.

The proposed CNN outperforms the first runner up in the Music Information Retrieval Evaluation eXchange (MIREX) 2016 and the winner of MIREX 2014 with a gain of $2.2702\mkern-2mu\sim\mkern-2mu 5.9563$ dB global normalized source to distortion ratio (GNSDR) when applied to the iKala dataset. An experiment with the DSD100 dataset on the full-tracks song evaluation task also shows that our model is able to compete with cutting-edge singing voice separation systems which use multi-channel modeling, data augmentation, and model blending. 


\end{abstract}

\section{Introduction}
\label{sec:Introduction}

Humans have an exceptional ability to separate different sounds from a musical signal~\cite{bregman1994auditory}. For instance, some musicians can distinguish the guitar part from a song and transcribe it; and most non-musician listeners are able to hear and sing along to lyrics of a song. Machines, however, have not yet mastered the ability to separate voices in music, despite the steep increase in the amount of research on artificial intelligence and music over the past few years~\cite{chuan2018, Herremans2017tax, van2013deep, ITASLP16:MultiChannel,ICASSP2017:DSD100Winner,ISMIR2017:UNet}. In this paper, we focus on the task of singing voice separation from a polyphonic musical piece, i.e., the automatic separation of a musical piece into two music signals: the singing voice and its music accompaniment. Some singing voice separation (SVS) systems~\cite{ITASLP12:MultiChannel,ICASSP15:MultiChannel,ITASLP16:MultiChannel,ICASSP2017:DSD100Winner} take this one step further by separating the music accompaniment into different types of musical instruments. In this research, we focus on the first task of separating the singing voice from its music accompaniment. The potential applications of automatic singing voice separation are plentiful, and include melody extraction/annotation~\cite{BigMM16:MIREX15Winner,ISMIR2017:AutoAnno}, singing skill evaluation~\cite{ICMC2014:iOS}, automatic lyrics recognition~\cite{mesaros2010automatic}, automatic lyrics alignment~\cite{wang2004lyrically}, singer identification~\cite{ICMLA2014:MFCC} and singing style visualization~\cite{ICMLA2014:Visual}. These applications are not only useful for researchers in the field of music information retrieval (MIR), but extend to commercial applications such as music for karaoke systems~\cite{wang2004lyrically}. 

We propose a novel convolutional neural network (CNN) approach for extracting a singing voice from its musical accompaniment. The key innovations in this design are the inclusion of Ideal Binary Mask (IBM)~\cite{Wang2005:IBM} as the target label, and the use of cross entropy~\cite{nielsen2015neural} as the training objective. This particular combination of IBM with cross entropy loss has proven to be extremely effective for image classification~\cite{VCPR2017:WeaklyLabel}. In the case of singing voice separation, the IBM represents a binary time$\times$frequency matrix, whereby a `1' indicates that the target energy is larger than the interference energy within the corresponding time-frequency (T-F) bin and `0' indicates otherwise. The training is guided by cross entropy, i.e., the average of the probability error between the predicted and the target label for each T-F bin. Additionally, we pretrain the weights of the CNN by training it as an autoencoder using singing voice spectrograms. The proposed network design enables us to leverage the power of CNNs for pixel-wise image classification, i.e., classifying each individual pixel of an image~\cite{NIPS2012:ImageNet,CVPR2015:FCN}. This is done performing multiclass classification (one class per sound source) for each T-F bin in our spectrogram, thus directly estimating the soft mask. This allows us to eliminate one of the very commonly used postprocessing step, the Wiener filter ~\cite{ITASLP12:MultiChannel,ISMIR2014:WorstDNN,ICASSP15:MultiChannel,ITASLP16:MultiChannel,BigMM16:MIREX15Winner,ICASSP2017:DSD100Winner,SVSGAN} (see Section~\ref{sec:RelatedWork}). 

We set up an experiment to test the proposed system with state-of-the-art models for SVS. When training our model on the iKala dataset~\cite{ICASSP2015:iKala}, we achieve $2.2702\mkern-2mu\sim\mkern-2mu 5.9563$ dB Global normalized source-to-distortion ratio (GNSDR) gain when compared to two state-of-the-art SVS systems~\cite{LVAICA2017:MIREX20161stRunnerUp,MIREX14:Winner}. 
A second experiment, on the full-track songs from the DSD100 dataset~\cite{SiSEC2016:DSD100}, shows no statistically significant difference between the proposed system and the current state-of-the-art systems. These experimental results suggest the need for a dataset agnostic model, meaning that instead of blindly feeding more data to models (which greatly improves training time), there is a need for efficient and effective models that perform well across different dataset, even with limited data. In the current research, we work towards this goal by using a network architecture that has shown to be effective in the field of image classification, and use a validation procedure during training and postprocessing to ensure that our CNN generalizes better. Furthermore, when designing our novel architecture, we trained and tested the model on two different datasets, such that the final optimized architecture would perform well across these datasets. 

In the next section, an overview of the current state-of-the art in voice separation models is given, followed by a description of our proposed CNN model with a formal definition of IBM and cross entropy. We then describe the details of the experimental setup and the training methodology, and present the results. Finally, conclusions regarding our proposed model and future research are offered.

\section{Related Work}
\label{sec:RelatedWork}

This section presents existing research in the field of singing voice separation. Experienced readers, who are familiar with the basics of the field, may skip to the sixth paragraph of this section for a detailed description of some of the latest state-of-the-art models. For a more comprehensive overview of the research undertaken in the last 50 years in this field, we refer the reader to the overview article~\cite{rafii2018overview}. 

The most popular preprocessing method in the field of singing voice separation involves transforming the time-domain signal into a spectrogram~\cite{ICMC2000:ISA,ISMIR2005:NMF,ITASLP2007:HarmPerc,ISAST2010:HarmPerc,ITASLP2010:AccSoundRed,ICASSP2012:rPCA,ISPL2014:HarmPerc,MIREX14:Winner}. Given that the value of each time-frequency (T-F) bin in the magnitude spectrogram~$X$ is non-negative, existing research on blind source separation (BSS) typically applies techniques such as Independent Subspace Analysis (ISA)\cite{ICMC2000:ISA} and Non-negative Matrix Factorization (NMF)~\cite{AINIPS2001:NMF}. The former, ISA, is a variant of Independent Component Analysis (ICA), which has previously been used to solve the cocktail party problem~\cite{cherry1953:Cocktail}. Independent Component Analysis is built upon the assumption that the number of mixture observation signals is equal to or greater than the target sources. The ISA variant, however, relaxes this constraint by using the non-negative spectrogram $X$. The second technique often used for blind source separation, NMF, decomposes $X$ into two non-negative matrices $L$ and $R$. The product of these two matrices approximates $X$, such that $LR \approx X$, with $D$ being the difference, such that $D = X - LR$. The matrix $D$ is later assumed to have the timbral characteristics of the singing voice.

NMF was the most widely adopted BSS technique in the 2000s~\cite{IJCNN2004:NMF,ISMIR2005:NMF,ITASLP2007:HarmPerc,NC2009:NMF,ISAST2010:HarmPerc,ISMIR2010:betaDiv}. The main difference between the various NMF-based methods is how the objective function is formulated. A typical formulation could be, $\min||X - LR||^2$ or $\min Div(X||LR)$, where $Div$ is the Kullback$\text{-}$Leibler divergence function. The popularity of NMF is partly due to the fact that the two matrices ($L$ and $R$) can easily be interpreted as a set of different types of musical instruments (or different tracks in the music), which we refer to as $I$. To understand this interpretation, let us first assume the columns of $L$ to be the frequency/tone basis functions $l_i$ and the rows of $R$ to be the time basis functions $r_i$, where $i$ is one of the musical instrument (or tracks) in the music. The factorized matrices ($L$ and $R$)  can be decomposed as the sum of the outer product of the basis functions, such that $LR = \sum_{i \in I} l_i\times r_i$. Thus, a frequency basis function $l_i$ can be interpreted as the timbre of instrument $i$. The corresponding set of time basis functions $r_i$ indicate how the sound of instrument $i$ evolves during the music. Additionally, $I$ is sometimes divided into two groups by posing constraints for the set of harmonic or pitched instruments (e.g. piano), $h \in I$, and the set of the percussion instruments (e.g. drum), $p \in I$~\cite{ITASLP2007:HarmPerc,ISAST2010:HarmPerc,ISPL2014:HarmPerc}.

A related technique, Robust Principal Component Analysis (rPCA), has also been applied to source sound separation~\cite{UIUCTech2010:rPAC}. It uses an augmented Lagrange multiplier to \textit{exactly}\footnote{NMF-based methods do not have this strong constraint. After their optimization process, it likely happens that the rank of $LR$ cannot be reduced to $|I|$, or that $D$ is not a sparse matrix.} separate $X$ into a low rank matrix and sparse matrix, $X = \sum_{i \in I} l_i\times r_i - D$, was widely adopted since 2012~\cite{ICASSP2012:rPCA}. The resulting factorized matrix $LR$ is a low rank approximation of $X$. The use of rPCA in source separation is motivated by the fact that (i) that the basis function of $LR$ approximates the spectrogram of the musical accompaniment component in the mixture signal; and (ii) $D$ is a sparse matrix that closely approximates the spectrogram of the separated singing voice. To better understand this, note that $X \approx LR$ and $X \approx \sum_{i \in I} l_i\times r_i$. If the number of musical instruments $|I|$ is the reduced rank of $X$, then $LR$ is a low rank approximation of $X$. Since the singing voice falls in between the harmonic instruments and percussion instruments, it is assumed to be represented by $D$.

\citet{MIREX14:Winner} use rPCA to obtain a sparse matrix, which is treated as a vocal time-frequency mask, and a vocal spectrogram. They then estimate the vocal F0 contour in this spectrogram in order to form a harmonic structure mask. By combining these two masks, they are able to better perform singing voice separation. This method, referred to as IIY, is the winner of MIREX 2014\footnote{\label{fn:MIREX2014}\url{http://www.music-ir.org/mirex/wiki/2014:Singing_Voice_Separation_Results}}. ~\citet{ICASSP2015:iKala} use the annotation of the vocal F0 contour to form a sparsity mask, which they then use as the input for rPCA to obtain a better vocal spectrogram. There exist several other approaches for source separation, such as the use of a similarity matrix~\cite{ICASSP2012:SimMix,ISMIR2012:SimMix}. Based on the MIREX 2014 results\footnotemark[\getrefnumber{fn:MIREX2014}], however, none of them outperform the rPCA-based methods. Hence, rPCA has become the de facto baseline in recent years.

Inspired by the influential work of~\citet{NIPS2012:ImageNet} on large-scale image classification from natural images, the use of deep learning has recently gained a lot of attention. Most deep-learning based SVS systems~\cite{ISMIR2014:WorstDNN,BigMM16:MIREX15Winner,ICASSP17:MIREX2016Winner,LVAICA2017:MIREX20161stRunnerUp,ICASSP2017:DSD100Winner} are trained to match the network input (i.e., the magnitude spectrogram of the mixture signal), with the target label (i.e., the ground truth magnitude spectrogram of the target sound source). Given enough training data, neural networks are typically able to estimate good approximations any continuous function \cite{NN1991:DeepAppro}, in this case, the magnitude spectrogram for each of the sound sources is estimated. These magnitude spectrograms, however, are not yet a good representation of the different sources. Contrary to intuition, these systems require a Wiener filter postprocessing step, in which a soft mask is calculated for the estimated magnitude spectrograms for every target sound source. These masks are then multiplied with the original magnitude spectrogram of the mixture signal to recreate each estimated signal. Using these soft masks typically gives a better separation quality than directly using the network output to synthesize the final signal~\cite{ICASSP2017:DSD100Winner}. This suggests that we should skip the Wiener filter postprocessing and design a network to learn a soft mask directly. 

Recent advances in the field of computer vision~\cite{CVPR2015:FCN} have greatly advanced image classification techniques by moving away from the image level towards the pixel-level. Pixel-wise classification aims at classifying each individual pixel in an image. The task of classifying each T-F bin of a spectrogram into a vocal or non-vocal component can be considered as a pixel-wise classification problem. 

Creating the pixel-wise ground truth for image segmentation typically involves extensive human effort. Luckily, this is not the case in SVS research as we can simply calculate the ground truth mask from a training set which contains the separated signals (see Section \ref{subsec:NetworkArch}). \citet{LVAICA15:IBM} and \citet{AES2016:IBM} perform singing voice separation using IBM as the target label for training a deep feed-forward neural network. 
In this research, however, we opt to use a convolutional neural network architecture, which has proven to greatly improve the performance of image classification tasks~\cite{NIPS2012:ImageNet,CVPR2015:FCN}. 
A similar CNN architecture for SVS, abbreviated in what follows as MC, has been proposed by \citet{LVAICA2017:MIREX20161stRunnerUp}. This method was the first runner up in the MIREX 2016 competition
\footnote{\label{fn:MIREX2016}\url{http://www.music-ir.org/mirex/wiki/2016:Singing_Voice_Separation_Results}}. The architecture proposed in this research improves the dimensions of the convolutional layer and introduces a cross entropy loss function, which greatly improves performance. 

Other state-of-the-art alternatives to using a CNN include the use of Recurrent Neural Networks (RNN)~\cite{ISMIR2014:WorstDNN} and bi-directional Long Short Term Memory (BLSTM) Networks~\cite{ICASSP2017:DSD100Winner}. These networks are designed to capture temporal changes, and may therefore not be necessary in a voice separation context. 

\citet{ISMIR2017:UNet} where the first to tackled SVS tasks by using a deep convolutional U-net in which the network predicts the soft mask. Their system shows remarkable performance on two datasets, iKala and MedleyDB~\cite{ISMIR2014:medleydb}. It should be noted, however, that while their network was tested on iKala and MedleyDB, it was trained on a gigantic dataset (the equivalent of two months worth of continuous audio) supplied by industry~\cite{humphrey2017mining}. This is much larger than the iKala and DSD100 training sets used in this research, which contain a total of respectively 76 minutes and 216 minutes of audio. The performance of similar U-net architectures~\cite{stoller2017adversarial,stoller2018jointly} trained on these smaller training set (e.g. DSD100) perform much worse than the original model. We can thus conclude that the remarkable performance reported by \citet{ISMIR2017:UNet} is mainly depended on the tremendous large training set, instead of the U-net architecture~\cite{humphrey2017mining}.

In this paper, we explore a CNN-based method with soft-mask prediction further improve the state-of-the-art in SVS systems. The next section will describe our proposed system in more detail. 

\section{CNN Network Design}
\label{sec:CNNDesign}

In this section, we first describe how the original mixture signal is transformed into a set of spectrogram excerpts, which are used as the input of the proposed CNN model. We then outline the network architecture, along with a formal definition of IBM and cross entropy. Next, we discuss issues related to the implementation and design of the CNN. Finally, an outline is given of how the network output is transformed into two separated signals, the singing voice and music accompaniment.

\subsection{Preprocessing}
\label{subsec:Pre-Processing}

In the preprocessing stage, the actual input for the CNN is created. First, we apply a Short-Time Fourier Transform (STFT) on the mixture signal $x$ to obtain the magnitude spectrogram $X$ and the phase spectrogram $pX$. For each Fast Fourier Transform (FFT) step, we use the Hann windowing function~\cite{Oppenheim:2009:DSP} with a window size $W$ of $46.44$ms, a hop size $H$ of $11.61$ms and a $4\plh$ zero padding factor. By setting the sampling rate $f_S$ at $22.05$ kHz, each FFT step is with size $N{=}4096$, $W{=}1024$ and $H{=}256$. This STFT configuration was chosen based on the authors' previous study on sinusoidal partials tracking~\cite{Interspeech2017:SinPart}. 

Sinusoidal partials tracking (PT) is a peak-continuation algorithm that links up the spectral peaks into a set of tracks. Each track models a time-varying sinusoid. The tracks are called partials when they represent the deterministic part of the audio signal. In the previous PT study, the average length of a singing voice partial was found to be around 9 continuous frames and the $4\times$ zero padding factor improved the separation quality of the ideal case. Hence we can assume that these settings should allow for enough temporal and spectral cues in order to properly train the CNN. The input of the proposed CNN consists of an image snapshot of $X$ with a shape of $(9\plh 2049)$, which is a spectrogram excerpt of $(9\plh256\plh1{,}000)/22{,}050=104.49$ms and $11.025$ kHz. 

\subsection{Network Architecture with Ideal Binary Mask and cross entropy}
\label{subsec:NetworkArch}

\begin{table}
\small
\begin{center}
    \caption{Network Architecture of the proposed CNN along with the configuration and the corresponding number of trainable parameters and features.}
    \label{tab:NetworkArch}
	\begin{tabular}{c|c|c}
    \toprule
\multirow{2}{*}{\textbf{Layer}}  & \multirow{2}{*}{\textbf{Configuration}}  & \textbf{Num. of} \\
& & \textbf{Trainable Parameters}\\
\midrule
\multirow{2}{*}{{Input}} & Input Size is $(9\plh2049)$ & \multirow{2}{*}{{N/A}} \\
& Num. of features is $(9\plh2049)=18,441$ & \\
\midrule
\multirow{2}{*}{{Convolution}} & $32@(3\plh12)$, Stride 1 & $(3\plh12)\plh32+32$ \\
 & Zero Pad, ReLU & $=1{,}184$ \\
\midrule
\multirow{2}{*}{{Convolution}} & $16@(3\plh12)$, Stride 1 & $(3\plh12)\plh32\plh16+16$ \\
& Zero Pad, ReLU & $=18{,}448$ \\
\midrule
\multirow{3}{*}{Max-Pooling} & Non-Overlap $(1\plh12)$ reshapes & \multirow{3}{*}{N/A} \\
& input size to $(9\plh12)=1{,}539$ & \\
& Num. of features is $(9\plh171)\plh16=24{,}624$ & \\
\midrule
\multirow{2}{*}{{Convolution}} & $64@(3\plh12)$, Stride 1 & $(3\plh12)\plh16\plh64+64$ \\
& Zero Pad, ReLU & $=36{,}928$ \\
\hline
\multirow{2}{*}{{Convolution}} & $32@(3\plh12)$, Stride 1 & $(3\plh12)\plh64\plh32+32$ \\
& Zero Pad, ReLU & $=73,760$ \\
\midrule
\multirow{3}{*}{Max-Pooling} & Non-Overlap $(1\plh12)$ reshapes & \multirow{3}{*}{N/A} \\
& input size to $(9\plh15)=135$ & \\
& Num. of features is $(9\plh15)\plh32=4{,}320$ & \\
\midrule
Dropout & with probability $0.5$ & N/A \\
\midrule
\multirow{2}{*}{{Fully-Connected}} & \multirow{2}{*}{$2{,}048$ Neurons, ReLU} & $4{,}320\plh2{,}048+2{,}048$ \\
& & $=8{,}849{,}408$\\
\midrule
Dropout & with probability $0.5$ & N/A \\
\midrule
\multirow{2}{*}{{Fully-Connected}} & \multirow{2}{*}{$512$ Neurons, ReLU} & $2{,}048\plh512+512$  \\
& & $=1{,}049{,}088$\\
\midrule
\multirow{3}{*}{Output} & $18{,}441$ Neurons, Sigmoid & $512\plh18{,}441+18{,}441$  \\
 & Reshape $(9\plh2049)$ Singing Voice & $=9{,}460{,}233$ \\
& IBM Label to match these Neurons & \\
\midrule
Objective Function & Cross Entropy & Total: $19,489,049$ \\
\bottomrule
	\end{tabular}
	\end{center}
\end{table}

Table \ref{tab:NetworkArch} shows the network architecture of the proposed CNN along with the configuration and the corresponding number of trainable parameters and features. We adopt the CNN architecture developed by~\citet{ISMIR2016:SalMap} for voice-detection. For that task, the network was trained on weakly labeled music\footnote{Each piece of music only has one annotation that indicates whether the music contains vocals or not.}. The resulting saliency map, created through guided backpropagation of the CNN, shows the singing voice in the T-F bin level. 

In the current research, we use the IBM as the target label instead of weak labels. IBM can be formally defined as follows. Let the $F\plh T$ matrix $X$ denote the magnitude spectrogram, whereby $F$ is the number of frequency bins, $F = (\floor[\big]{\frac{N}{2}}+1)$ with $N$ as the FFT size, and $T$ is the number of frames. Given the magnitude spectrogram of the voice $X_V$ and of the music accompaniment $X_S$, the IBM of the singing voice, which is a $F\plh T$ matrix $B$, is calculated as,
\begin{equation}
	B[n,t] = 
\begin{dcases}
    1, & \text{if } X_V[n,t] > X_S[n,t]\\
    0, & \text{otherwise}
\end{dcases}
\label{eq:IBM}
\end{equation}
where $t\in[1,T]$ is the time index and $n\in[1,F]$ is the frequency bin index. The IBM of the music accompaniment is denoted as $\overline{B} = \abs{1-B}$. 

The resulting matrix $B$ forms the target label of the neural network. Together with the network predictions, $Y[n,t]$, formed by the sigmoid output of the final layer, we can calculate the cross entropy over all T-F bins, as:
\begin{equation}
\begin{aligned}
	C[n,t] = & B[n,t]\plh-log(Y[n,t]) + \\
     		 & (1-B[n,t])\plh-log(1 - Y[n,t])\\
\end{aligned}
\label{eq:CrossEntropyFTbin}
\end{equation}

The training objective of our proposed network minimizes the cross entropy. This type of objective function performs better then that often used softmax function, as it is tailored to the fact that each T-F bin can have multiple labels. Unlike a pixel in an image whose value is paired with the desired label, the value of a T-F bin in the magnitude spectrogram of a mixture signal is roughly the sum of the T-F bin of the singing voice and its accompaniment. 

Alternative training objectives were explored, such as minimum mean square error (MMSE) with both IBM and Ideal Ratio Mask (IRM)~\cite{wang2014training} as the target label. We found, however, that the MMSE does not decrease much with IRM and IBM; and that cross entropy also does not decrease much with IRM. We therefore opted to integrate IBM with a cross entropy training objective. 

To improve the network performance, the weights were first initialized with Xavier's initializer~\cite{AISTATS2010:XavierInitial}. To further improve these initial weights, the CNN trained as an autoencoder using spectrogram excerpts of the ideal singing voice for 300 epochs. These initial weights allow us to train the resulting separation network much more efficiently. 

An often used technique to speed up a model's convergence is Batch Normalization (BN)~\cite{ICML2015:BatchNorm}. This technique requires a number of extra parameters, and increases the training time for each epoch. When implementing BN in our network, we did not notice an improvement in training time, and most importantly, there was no improvement of the separation quality. We therefore opted not to include BN in the proposed system. Similarly, we also did not find an improvement of separation quality and training time when we used the skip connection method~\cite{CVPR2017:DenseNetwork} and the method of converting the fully-connected layer to a convolutional layer~\cite{CVPR2015:FCN}. Hence, both methods were not included in the proposed CNN.

Existing network architectures commonly apply a ($3\plh3$) filter in the convolutional layers. Because we applied $4\plh$ zero padding factor in the frequency domain during the STFT calculation, we set the convolutional filter size to be ($3\plh12$), whereby 3 represents the time and 12 the frequency bin. The time dimension in the pooling layer was not reduced as this can introduce jitter and other artifacts. The frequency dimension in the max pooling layer, however, was reduced. This process is roughly analogous to Mel-frequency calculation, which has been empirically proven to provide useful features for audio classification tasks~\cite{ICMC2008:MFCC,ESPC10:MFCC,ISMIR2011:SingVoiceDetect}. The number of features maps in each convolutional layer is halved compared to the original voice-detection CNN architecture~\cite{ISMIR2016:SalMap}, so as to shorten the training time, and most importantly, to avoid degradation of the separation quality. Finally, the dropout~\cite{JMLR2014:Dropout} settings and ReLU activations~\cite{NIPS2012:ImageNet} are preserved as in the original architecture. 

\subsection{Postprocessing}
\label{subsec:postprocessing}

The goal of the singing voice separation task is to get two isolated music signals: voice and accompaniment. We therefore need to convert the estimated soft mask by network into two audio signals. In order to do this, the CNN output is first reshaped from $(1\plh 18{,}441)$ to $(9\plh 2{,}049)$ in order to reconstruct the 9 frames. The estimated network output, before postprocessing, is considered to be the \emph{soft mask} of the estimated singing voice spectrogram, meaning that the value for each T-F can range from 0 to 1. This assumption is justified by the fact that IBM was selected as the target label during training and thus used to calculate the cross entropy with sigmoid function. The value of each T-F bin in the soft mask can be interpreted as the probability $e$ that the T-F bin belongs to the singing voice. 

To further improve the separation quality, we carry out the following optional refinement using the validation set. For a threshold $\theta$, we set $e$ to zero when $e<\theta$. Based on an experiment using the validation set (see Section \ref{sec:ExpSetup}), we set $\theta$ to be $0.35$ for the iKala dataset and $0.15$ for the DSD100 dataset.

\begin{figure}[h!]
  \centering
  \includegraphics[width=1\columnwidth]{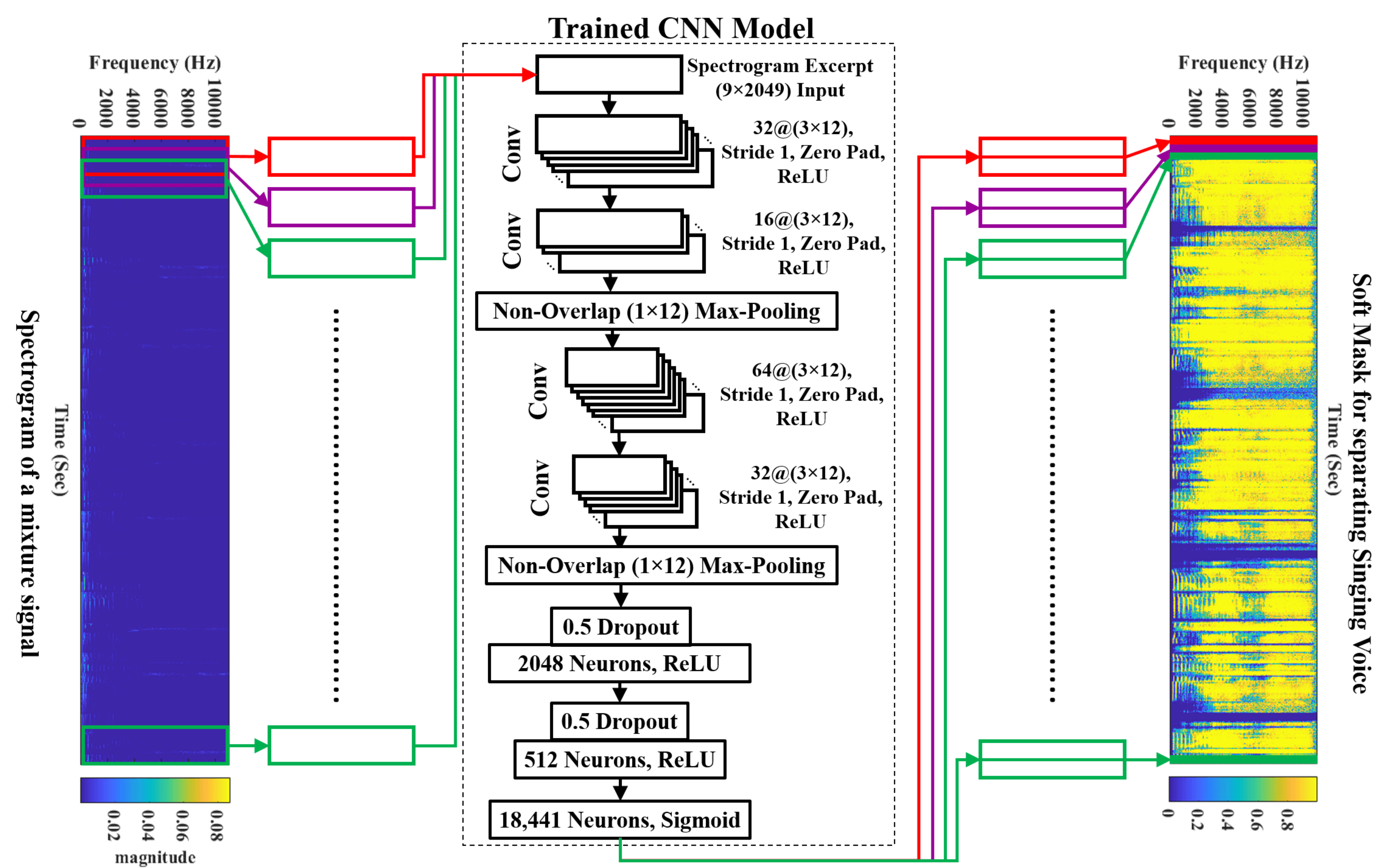}
  \caption{Architecture for estimating a soft mask based on an entire track.}
  \label{fig:softmask}
\end{figure}

The neural network architecture described above takes 9 audio frames as input. In order to estimate a single soft mask $M_V$ for separating the singing voice from an \emph{entire} song, we follow a two step approach inspired by~\citet{ISMIR2016:SalMap}. First, overlapping spectrogram excerpts (each 9 frames long) are fed into the network with a hop size of 1 frame. The middle frames of each estimated soft mask is then concatenated to create $M_V$. These two steps are illustrated in Figure \ref{fig:softmask}. The soft mask $M_S$ for obtaining the \emph{music accompaniment} from a test song can be calculated by $1 - M_V$.

Finally, the isolated signing voice signal is obtained by calculating the inverse TFT (iSTFT) of the element-wise multiplication between the estimated $M_V$ and $X$, and the original phase spectrogram $pX$. Similarly, we can obtain the isolated musical accompaniment signal by calculating the iSTFT of the element-wise multiplication between $M_S$ and $X$ using $pX$. In the case of a stereo recording, all of the procedures mentioned above should be carried out for each channel separately.

\section{Experiment Setup}
\label{sec:ExpSetup}

The separation quality of the proposed CNN model is evaluated and compared to other state-of-the-art SVS systems. This is achieved by using two datasets that are specifically designed for the SVS task. 
Before discussing the results of our experiment in the next section, a brief description of the music clips in each dataset is given, together with how these are divided into development and test sets. We then describe the evaluation procedure and discuss how the proposed CNN should be properly trained, so that a state-of-the-art results can be obtained.

\subsection{iKala Dataset}
\label{subsec:iKalaDataset}

The iKala dataset~\cite{ICASSP2015:iKala} is a public dataset specifically created for the SVS task. Each clip in the dataset is recorded in a CD quality wave file and sampled at $44.1$ kHz, with two channels. One channel consists of the ground truth singing voice $V$, and the other one forms the ground truth music accompaniment $S$. The mixture signal $M$ is simply the sum of $V$ and $S$. There are 6 singers, of which three were female and three male. The singing voice tracks were almost entirely performed by one or more of these singers. The musical accompaniment tracks were all performed by professional musicians. Each clip is $30$ sec long and contains non-vocal regions with varied duration. The language of the lyrics is either English, Mandarin, Ksorean, or Taiwanese. The dataset contains 352 music clips, 100 of them are reserved for the evaluation of the MIREX\footnote{\url{http://www.music-ir.org/mirex/wiki/MIREX_HOME}} singing voice separation task and are not publicly available. Among the remaining 252 clips, 137 of these clips are labeled \textit{Verse} and 115 clips as \textit{Chorus}. 

In order to properly evaluate our proposed model, the 252 music clips in the iKala dataset were randomly divided into 3 sets, namely training, validation, and test set. The training set consisted of 152($\sim60\%$) clips, 50 ($\sim20\%$) music clips form the validation set and 50 ($\sim20\%$) the test set. The details of each set are described in Table~\ref{tab:TrainValidTestSets}.

\begin{table}[h!]
	\setlength\tabcolsep{2pt}
    \begin{center}
    \caption{The training, validation and test set split based on the iKala dataset. The numbers represent the file name of the corresponding wave file.}
	\label{tab:TrainValidTestSets}
	\begin{tabular}{ll|l|c}
\toprule
 & \multicolumn{2}{c}{\textbf{Music Clips}} & \textbf{Total} \\
 & \multicolumn{1}{c}{\textbf{Verse}} & \multicolumn{1}{c}{\textbf{Chorus}} & \textbf{Clips} \\
\midrule
\multicolumn{1}{c}{\multirow{16}{*}{\textbf{Training}}} & 10174, 21025, 21031, 21032, 21033, & 10171, 10174, 21033, 21035, 21038, &  \multicolumn{1}{c}{\multirow{16}{*}{152}} \\
\multicolumn{1}{c}{} & 21035, 21038, 21039, 21040, 21054, & 21040, 21054, 21056, 21057, 21059, & \\
\multicolumn{1}{c}{} & 21055, 21059, 21060, 21063, 21064, & 21061, 21063, 21068, 21074, 21075, & \\
\multicolumn{1}{c}{} & 21069, 21076, 21086, 31081, 31099, & 21083, 21086, 31047, 31075, 31083, & \\
\multicolumn{1}{c}{} & 31101, 31104, 31107, 31109, 31113, & 31101, 31103, 31112, 31113, 31115, & \\ 
\multicolumn{1}{c}{} & 31114, 31119, 31134, 31136, 31143, & 31118, 31135, 45305, 45358, 45361, & \\ 
\multicolumn{1}{c}{} & 45305, 45358, 45359, 45362, 45367, & 45363, 45367, 45368, 45369, 45378, & \\ 
\multicolumn{1}{c}{} & 45368, 45378, 45381, 45382, 45386, & 45382, 45384, 45386, 45387, 45392, & \\
\multicolumn{1}{c}{} & 45387, 45388, 45389, 45390, 45393, & 45398, 45406, 45413, 45422, 45424, & \\
\multicolumn{1}{c}{} & 45398, 45404, 45414, 45415, 45421, & 45425, 45428, 45429, 54189, 54190, & \\
\multicolumn{1}{c}{} & 45423, 45428, 45429, 45434, 54173, & 54192, 54202, 54211, 54220, 54221, & \\
\multicolumn{1}{c}{} & 54186, 54191, 54192, 54194, 54205, & 54223, 54226, 54233, 54236, 54239, & \\
\multicolumn{1}{c}{} & 54223, 54226, 54245, 54246, 61670, & 54243, 54245, 54246, 54249, 61647, & \\
\multicolumn{1}{c}{} & 61671, 61673, 61674, 66558, 66564, & 61671, 61676, 61677, 66556, 66557, & \\
\multicolumn{1}{c}{} & 66565, 71706, 71710, 71711, 71719, & 71710, 71716, 71719, 71720, 71726, & \\ 
\multicolumn{1}{c}{} & 80612 & 90586 & \\ 
\midrule
\multicolumn{1}{c}{\multirow{5}{*}{\textbf{Validation}}} & 10161, 10171, 21068, 31092, 31129, & 10170, 21025, 21045, 21073, 21084, & \multicolumn{1}{c}{\multirow{5}{*}{50}} \\
\multicolumn{1}{c}{} & 31139, 31142, 45369, 45384, 45400, & 31092, 31100, 31129, 31137, 31143, & \\
\multicolumn{1}{c}{} & 45409, 45417, 45422, 45435, 54016, & 45381, 45385, 45389, 45416, 45419, & \\ 
\multicolumn{1}{c}{} & 54189, 54219, 54242, 66559, 66560, & 45435, 54173, 54183, 54210, 54212, & \\ 
\multicolumn{1}{c}{} & 66563, 66566, 71712, 71720, 90586  & 54228, 66559, 66561, 66563, 71711  & \\ \midrule
\multicolumn{1}{c}{\multirow{8}{*}{\textbf{Test}}} & 21045, 21058, 21061, 21062, 21071, & 10161, 10164, 21058, 31093, 31109, & \multicolumn{1}{c}{\multirow{8}{*}{50}}\\
\multicolumn{1}{c}{} & 21073, 21075, 21084, 31083, 31117, & 31116, 31126, 31134, 31139, 45412, & \\
\multicolumn{1}{c}{} & 31132, 31135, 31137, 31144, 45391, & 45415, 54194, 54213, 54227 & \\ 
\multicolumn{1}{c}{} & 45392, 45410, 45412, 45416, 45418, & & \\ 
\multicolumn{1}{c}{} & 45431, 54190, 54213, 54216, 54227, & & \\
\multicolumn{1}{c}{} & 54233, 54243, 54247, 54249, 54251, & & \\
\multicolumn{1}{c}{} & 61647, 66556, 71723, 80614, 80616, & & \\
\multicolumn{1}{c}{} & 90587 & & \\
 \bottomrule
	\end{tabular}
    \end{center}
\end{table}

\subsection{Evaluation under iKala Dataset}
\label{subsec:evaliKala}

In line with the MIREX2016 evaluation procedures, we use a standard quality assessment tool for evaluating SVS systems called BSS Eval Version 3.0~\cite{ITASLP2006:BSS}. For each estimated/original clip, four quality metrics are calculated in order to assess the separation quality, namely Source to Distortion Ratio (SDR), source Image to Spatial distortion Ratio (ISR), Source to Interferences Ratio (SIR), and Sources to Artifacts Ratio (SAR). The global separation quality for each clip in terms of singing voice, is measured by the normalized SDR (NSDR). This ratio is calculated as

\begin{equation}
	\text{NSDR}(\overline{V},V,M) = \text{SDR}(\overline{V},V) - \text{SDR}(M,V)
	\label{eq:NSDR}
\end{equation}

Here, $\overline{V}$ represents the audio signal of the estimated singing voice. The overall singing voice separation quality on a test set is determined by the global NSDR (GNSDR). This ratio is calculated as 

\begin{equation}
	\text{GNSDR} = \frac{1}{|\Lambda|}\sum_{i \in \Lambda}\text{NSDR}(\overline{V_i},V_i,M_i)
	\label{eq:GNSDR}
\end{equation}

whereby $\Lambda$ is a set of test clips; and the total number of the test clips is represented by $|\Lambda|$. A better separation quality is reflected by a larger GNSDR. Similarly to the quality of the singing voice, the above formula can be modified to calculate the separation quality of the music accompaniment by replacing $V$ by $S$ and $\overline{V}$ by $\overline{S}$ respectively. The GNSDR calculation is computationally expensive, hence we used parallel processing through a GPU\footnotemark[\getrefnumber{fn:GPU}] to accelerate this process. 

\subsection{DSD100 Dataset}
\label{subsec:DSD100Dataset}

The DSD100 dataset~\cite{SiSEC2016:DSD100} is a public dataset, specifically created for evaluating source separation algorithms capable of separating professionally produced music recordings into either two stereo signals (i.e., music accompaniment and singing voice), or five stereo signals (i.e., singing voice, music accompaniment, drums, bass and other). There are four wave files for each recording, in addition to the mixed recording wave file: the ground truth singing voice $V$, drums $U$, bass $A$ and other $O$. The ground truth music accompaniment $S$ is simply the sum of $U$, $A$ and $O$. The mixture signal $M$ is the sum of $V$ and $S$. The recordings are all in English, and feature different artists and genres. For example, the genres includes Rap, Rock, Heavy Metal, Pop and Country. The time duration ranges from $2$ min and $22$ sec to $7$ min and $20$ sec, with an average duration of $4$ min and $10$ sec. There are 100 recordings, that are evenly distributed over the development (dev) set and the test set. We used the dev set to create the training and validation set by following the procedures described in Section~\ref{subsec:training}.

\subsection{Evaluation under DSD100 Dataset}
\label{subsec:evalDSD100}

To enable easy comparison with other algorithms, we follow the evaluation procedure of the SiSEC 2016 MUS track, and use BSS Eval Version 3.0~\cite{ITASLP2006:BSS} to assess the separation quality of our SVS algorithm. In order to assess the separation quality of whole songs, however, we carry out the procedures below instead. 

The stereo mixture signal of each recording is first divided into a set of $30$ sec long music clips with $15$ sec overlap. We then exclude music clips which are smaller than $30$ sec or yield NaN (Not a Number) SDR values for the singing voice. The NaN SDR values mostly occur at the beginning and end of the recording, where there is no singing voice. 

We refer to the set of $30$ sec long clips for a recording $r$ as $\Lambda_r$. In order to assess the singing voice separation quality of a SVS algorithm, we first calculate the representative $(\text{SDR}_r)$ value of a recording $r$ by averaging the singing voice SDR for each clip $i$ in $r$, such that $SDR_r = \frac{1}{|\Lambda_r|}\sum_{i \in \Lambda_r}\text{SDR}(i)$. The singing voice separation quality of a SVS algorithm is represented by the median of these $SDR_r$ over the test set. The separation quality of other sound sources can be calculated similarly.

\subsection{Training}
\label{subsec:training}

The training instances were created by dividing each training song into a set of $(9\plh 2{,}049)$ spectrogram excerpts (one spectragram for each 9 frames) using a hop size of 8 frames ($92.88$ms). Since there is an overlap of only 1 frame, the training instances are concise. In the case of stereo recordings, each channel was processed in the same manner, but we chose to alternatingly use the spectrogram excerpts from one or the other channel, in order to have the same number of training instances as for the single channel. This procedure reduces the number of training instance significantly, yet preserve most of the information of each channel. Both datasets are evaluated on the basis of $30$ sec music clips. Using our network setup, a $30$ sec music clips equates to $30\plh1000/92.88 = 323$ input slices. For the ikala dataset, there are 152 clips of $30$ sec, resulting in $323\times 152 = 49{,}096$ training instances. For the DSD100 dataset, there are 347 clips of each $30$ sec, resulting in $323\times 347 = 112{,}081$ training instances. For each clip, we randomly shuffle the training instances for the purpose of regularization. In a similar fashion, validation instances are created using the set of validation songs. They are used for parameter initialization and model selection.

We use the Tensorflow~\cite{tensorflow2015-whitepaper} version of the ADAM~\cite{arXiv2014:adam} optimizer with its default values, to train a CNN for each dataset. The network is updated per batch of 171 instances. A BizonBox\footnote{\label{fn:GPU}\url{https://bizon-tech.com/}} with NVIDIA GTX TITAN X was used to train both CNNs. Each training epoch needed around $2$ min and $6$ min for the iKala and DSD100 dataset respectively. For regularization purposes, we used $50\%$ dropout~\cite{JMLR2014:Dropout} and shuffled the training instances. The target values were set to 0.02 and 0.98 instead of 0 and 1, as suggested by~\citet{ISMIR2016:SalMap}. This method prevents overfitting more so than L2 weight regularization. 

\begin{figure}[h]
	\centering
	\subfigure[The loss function for iKala Dataset]{\label{fig:CrossEntropyA}\includegraphics[width=60mm]{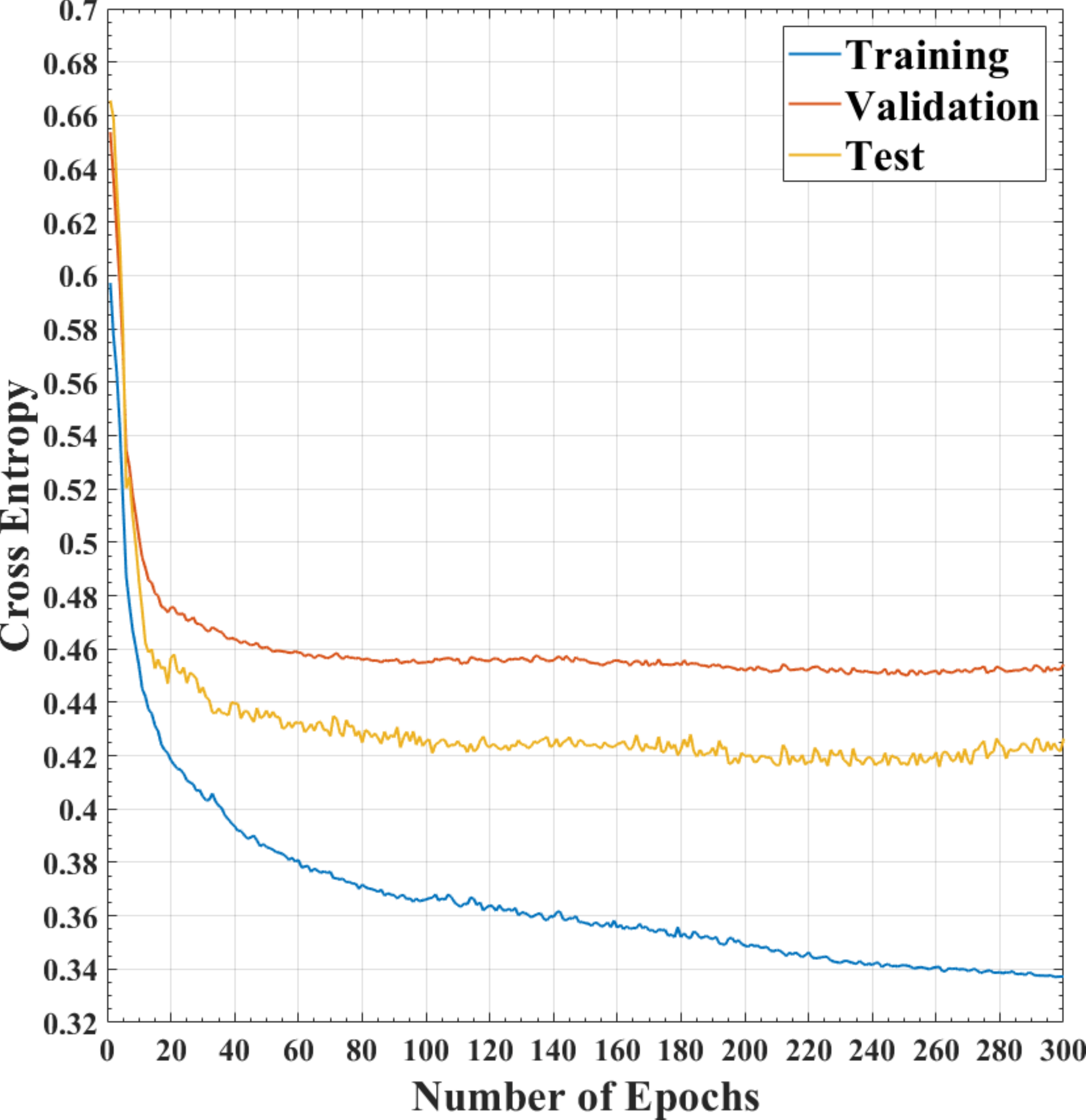}}
	\subfigure[The loss function for DSD100 Dataset]{\label{fig:CrossEntropyB}\includegraphics[width=60mm]{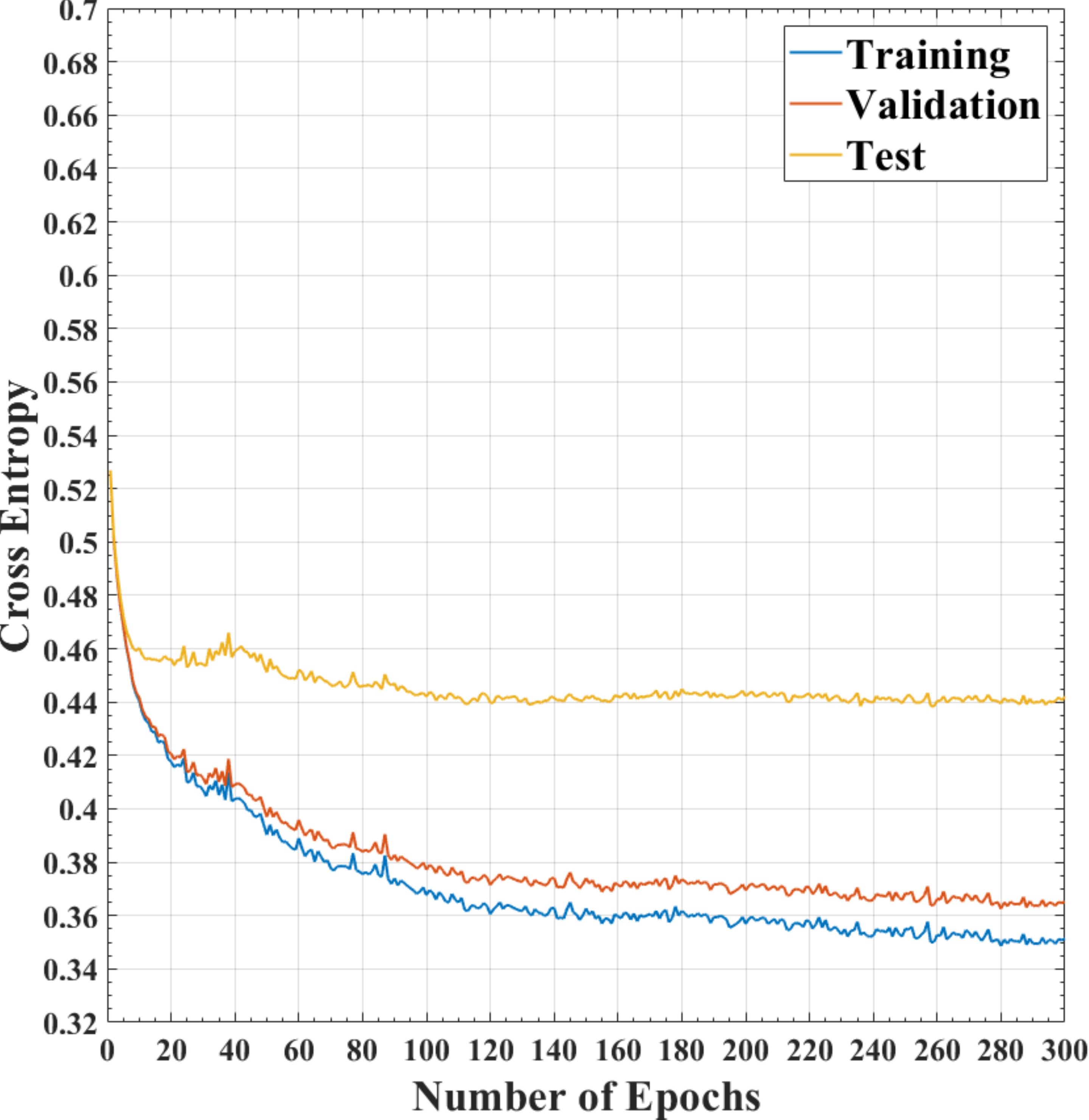}}
	\caption{Evolution of the cross entropy loss for each dataset during training. The lowest cross entropy loss of the validation set is $0.4509$ and $0.3625$ for the iKala and DSD100 dataset respectively. The final selected model for the iKala and DSD100 dataset was trained with 242 epochs and 280 epochs respectively.}
    \label{fig:CrossEntropy}
\end{figure}

All trainable parameters in our CNN were initialized with Xavier's initializer~\cite{AISTATS2010:XavierInitial}. In order to even further improve the set of initial parameters for the SVS task, the CNN is first treated as an \emph{auto-encoder} by pre-training it with spectrogram excerpts of the ideal singing voice for 300 epochs. The model with the lowest cross entropy loss for the validation set is then selected as the initial model for the actual training with the full network. After this parameter initialization, the proposed CNN is trained by feeding it the spectrogram excerpts of the mixture signal and the corresponding singing voice IBM as the target label. Figure~\ref{fig:CrossEntropy} shows the evolution of the cross entropy loss for each dataset. Note that we also plot the cross entropy loss of the test set for the sake of completeness. The final model is selected based on the lowest cross entropy loss on the validation set, which is $0.4509$ and $0.3625$, for the iKala and DSD100 dataset respectively. The selected model for the iKala and DSD100 dataset are trained with 242 epochs and 280 epochs respectively in order to ensure that the validation set has the lowest cost. The separation quality results of these models on the test set are described in the next section. 

\section{Experimental Results}

\begin{figure}[h!]
  \centering
  \includegraphics[width=1\columnwidth]{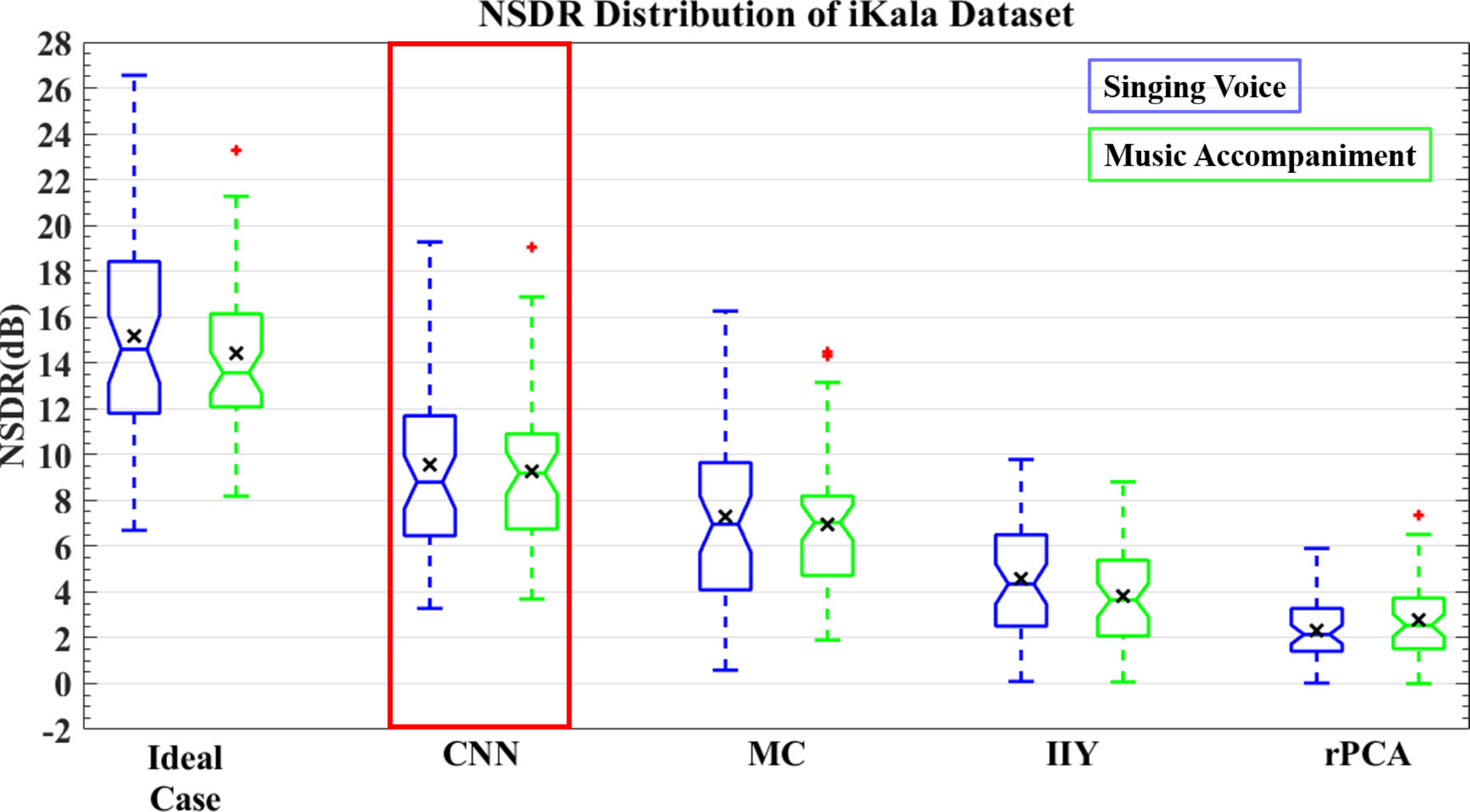}
  \caption{The NSDRs distribution of each SVS algorithm. The marks x indicate the GNSDRs of each SVS algorithm. The left bar represents the ideal GNSDR: $15.1944$ dB for singing voice, and $14.4359$ dB for musical accompaniment.}
  \label{fig:CompEavl}
\end{figure}

Using the \textbf{iKala dataset}, the proposed CNN was compared with the first runner up (MC) of MIREX 2016~\cite{LVAICA2017:MIREX20161stRunnerUp}, the winner (IIY) of MIREX 2014~\cite{MIREX14:Winner} and the rPCA baseline~\cite{ICASSP2012:rPCA}. A comparison of our model with the winner of MIREX 2016~\cite{ICASSP17:MIREX2016Winner} and MIREX 2015~\cite{BigMM16:MIREX15Winner} was not possible, as both winners do not share sufficient information to ensure a fair comparison. For example, they do not share their trained model, information on the training set, nor their separation results for each music clip\footnote{The 2016 winner~\cite{ICASSP17:MIREX2016Winner} has created a web service for others to try their separation method, however, each separated clip is only 10 sec long.}. The results\footnote{\label{fn:results}Readers who are interested in other evaluation metrics of our CNN model, may refer to \url{https://kinwahedwardlin.wordpress.com}} of our experiment are displayed in Figure~\ref{fig:CompEavl}. The CNN proposed in this paper achieves the highest GNSDRs for both singing voice and music accompaniment: $9.5774$ dB and $9.2484$ dB respectively. For the singing voice, our system achieves $2.2702$ dB higher than MC, $5.0908$ dB higher than IIY, and $5.9071$ dB higher than rPCA. For the music accompaniment voice, the proposed CNN achieves $2.3804$ dB higher than MC, $5.9563$ dB higher than IIY, and $6.5947$ dB higher than rPCA. To further justify that our CNN outperforms the others, we perform a one-way ANOVA, the results of which are summarized in Table \ref{tab:ANOVA_Comp}. The $p$-values confirm that the proposed CNN achieves a statistically significant GNSDR difference ($<$~0.01) compared to the other systems.

\begin{table}[h!]
	\begin{center}
        \caption{The significant GNSDR difference between each pair of the SVS systems evaluated by a One-way ANOVA test.}
	\begin{tabular}{l|cc|cc}
	\toprule
	\multirow{2}{*}{\textbf{Pair}} & \multicolumn{2}{c}{\textbf{Singing Voice}} & \multicolumn{2}{|c}{\textbf{Music Accompaniment}} \\
   & \textbf{F(1,98)}  & \textbf{$p$-value} & \textbf{F(1,98)}  & \textbf{$p$-value} \\
\midrule
CNN, MC & \small$8.4989$ & \small$0.0044$ & \small$9.2806$ & \small$0.0002$ \\
CNN, IIY  & \small$57.9684$  & \small$1.676\times 10^{-11}$ & \small$76.0115$ & \small$9.7516\times 10^{-16}$ \\
CNN, rPCA  & \small$59.7874$ & \small$9.4109\times 10^{-12}$ & \small$147.3874$ & \small$3.0223\times 10^{-21}$ \\
MC, IIY  & \small$17.9755$ & \small$5.0706\times 10^{-5}$ & \small$35.8675$ & \small$3.4918\times 10^{-8}$ \\
MC, rPCA  & \small$22.838$ & \small$6.1939\times 10^{-6}$ & \small$66.96450$ & \small$1.0299\times 10^{-12}$ \\
IIY, rPCA  & \small$1.5871$ & \small$0.2107$ & \small$1.5620$ & \small$0.2143$ \\
\bottomrule
	\end{tabular}
	\label{tab:ANOVA_Comp}
	\end{center}
\end{table}

\begin{figure}[h!]
	\centering
	\subfigure[Singing Voice]{\label{fig:DSDResultVoice}\includegraphics[width=115mm,height=50mm]{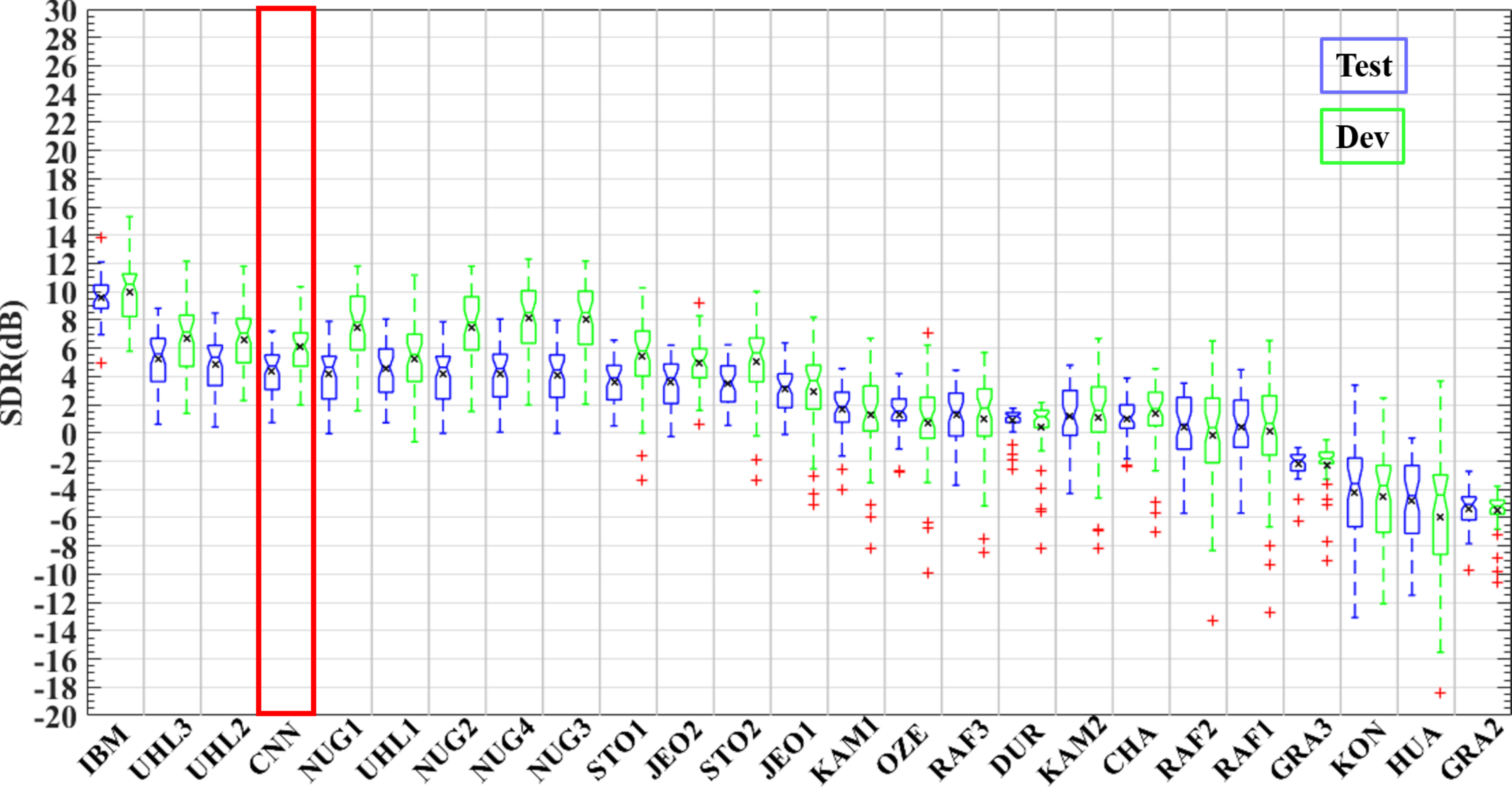}}
	\subfigure[Music Accompaniment]{\label{fig:DSDResultSong}\includegraphics[width=115mm,height=50mm]{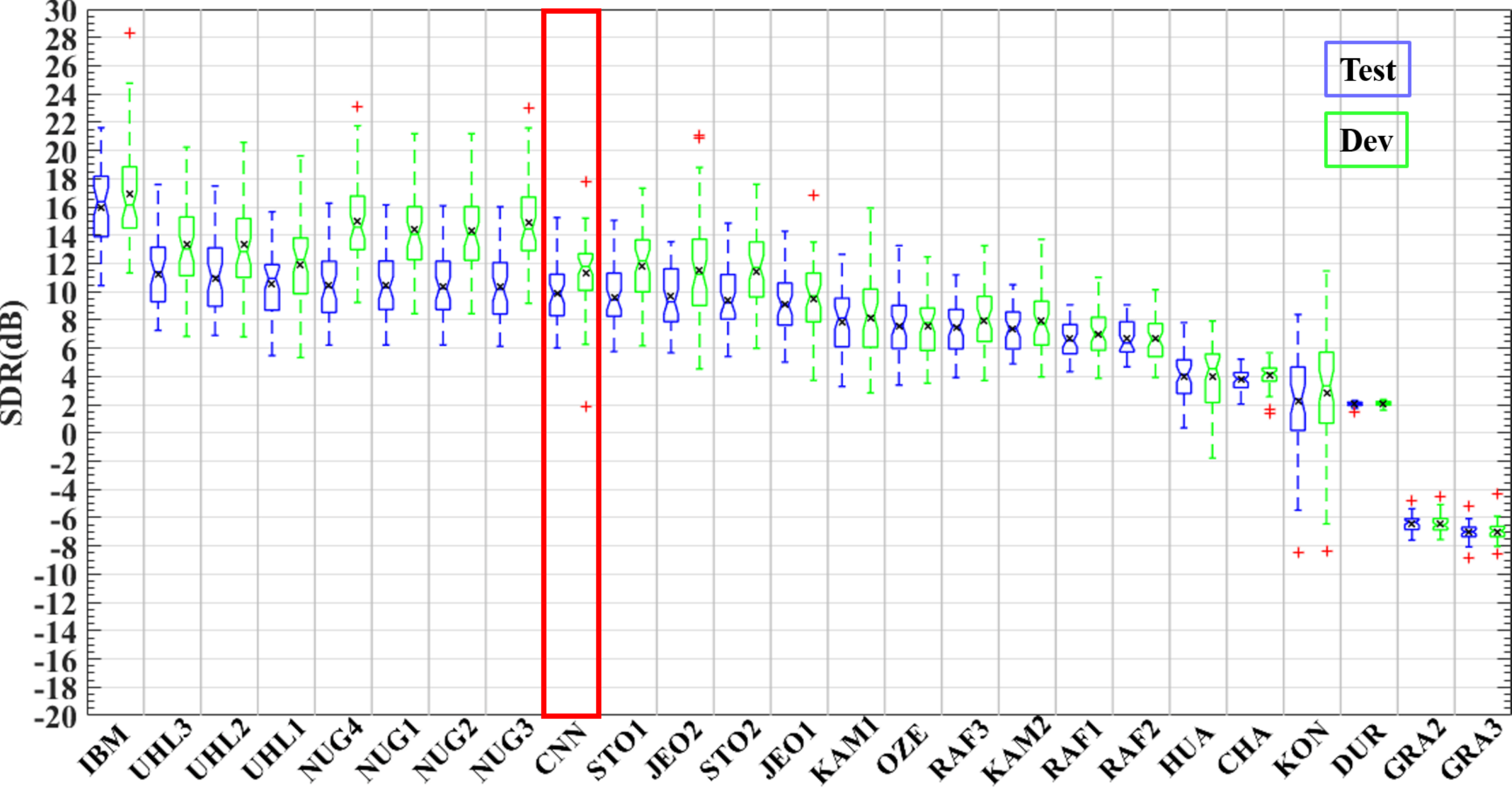}}
	\caption{The SDR distribution for the dev and test set, sorted by the median values of the test set for all SVS algorithms. For the Test set, our CNN achieves $4.7385$ dB and $9.8567$ dB for the singing voice and its accompaniment respectively. For Dev set, our CNN achieves $6.1632$ dB and $11.7888$ dB for the singing voice and its accompaniment respectively.}
    \label{fig:DSDResult}
\end{figure}

Secondly, the \textbf{DSD100 dataset} was used to compare the proposed CNN to the SVS systems that participated in the SiSEC 2016 MUS track\footnote{\label{fn:SiSEC2016}\url{http://sisec17.audiolabs-erlangen.de/}}. This track included 10 blind source separation methods: CHA~\cite{LVAICA2017:MIREX20161stRunnerUp}, DUR~\cite{IJSTAP11:DUR}, KAM~\cite{ICASSP15:KAM}, OZE~\cite{ITASLP12:MultiChannel}, RAF~\cite{ICASSP2012:SimMix,ISMIR2012:SimMix,ITASLP13:RAF}, HUA~\cite{ICASSP2012:rPCA} and JEO~\cite{LVAICA17:JEO}, and 14 supervised learning methods, which use different types of deep neural networks, including GRA~\cite{AES2016:IBM}, KON~\cite{ITASLP15:KON}, UHL~\cite{ICASSP2017:DSD100Winner}, NUG~\cite{ITASLP16:MultiChannel}, STO~\cite{ICASSP16:STO} and their variants, e.g. UHL1 and UHL2. Given the published details of their separation results\footnote{\url{https://github.com/faroit/sisec-mus-results}}, we are able to show the SDR distribution\footnotemark[\getrefnumber{fn:results}] for each SVS algorithm in Figure~\ref{fig:DSDResult}. Based on the median values for each clip in the test set, the proposed CNN ranks 3rd and 8th in term of the separation quality of the singing voice and the music accompaniment respectively. Its performance is just behind UHL and NUG which use multi-channel modeling~\cite{ITASLP16:MultiChannel}, data augmentation~\cite{ICASSP2017:DSD100Winner}, and model blending~\cite{ICASSP2017:DSD100Winner}. When interpreting these results, one should keep in mind that we only used $1\times10^5$ training instances to train the CNN (without data augmentation), whereas UHL was trained on $2\times10^6$ instances. This further illustrates the effectiveness of our network design. The result also shows that our proposed way of proprocessing training instances effectively reduces the size of the required training set. Furthermore, unlike the UHL1 model, our model does not require us to train a model separately for each channel.

\begin{figure}[h!]
	\centering
	\subfigure[Singing Voice]{\label{fig:DSDPValueVoice}\includegraphics[width=60mm,height=60mm]{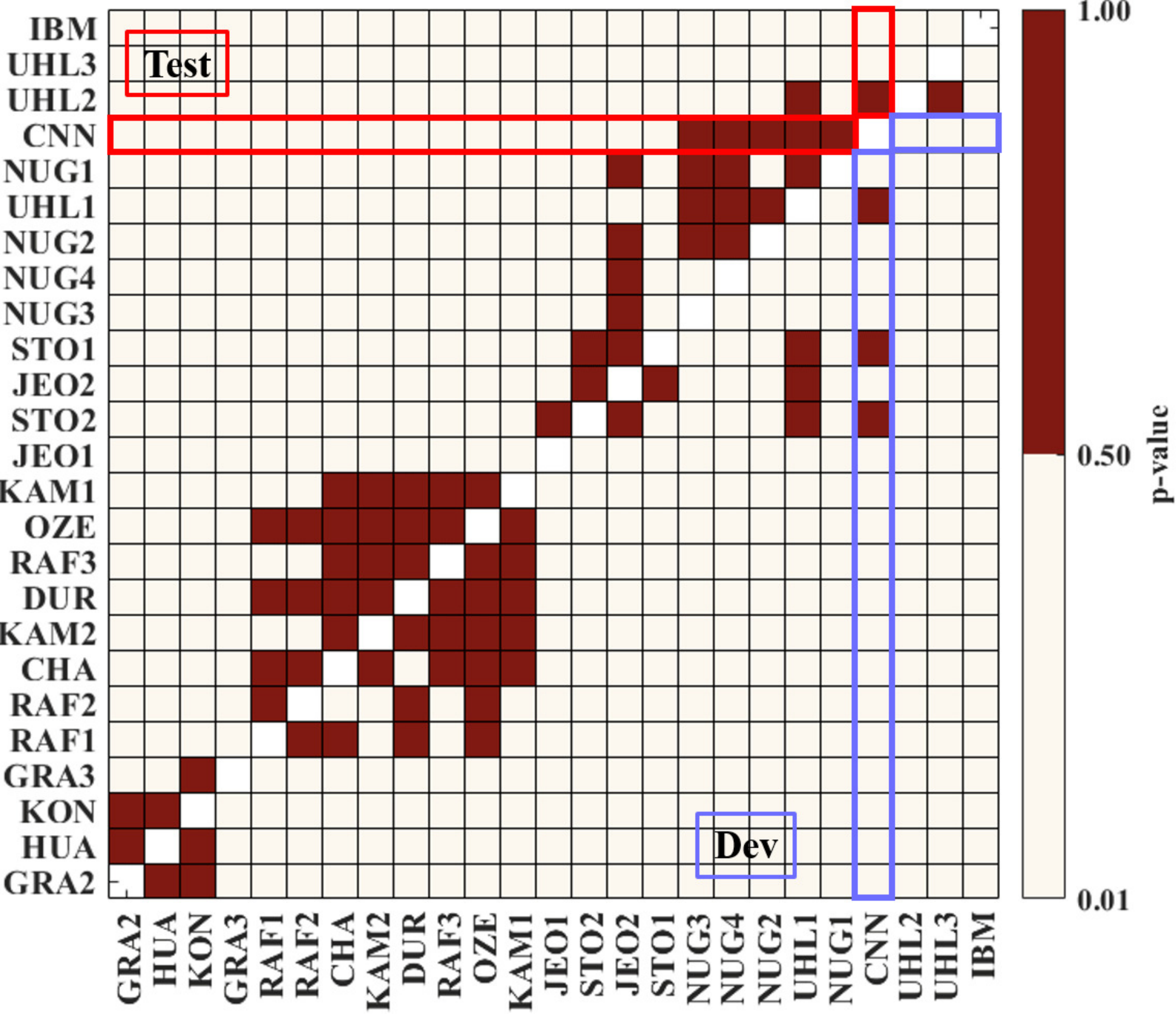}}
	\subfigure[Music Accompaniment]{\label{fig:DSDPValueSong}\includegraphics[width=60mm,height=60mm]{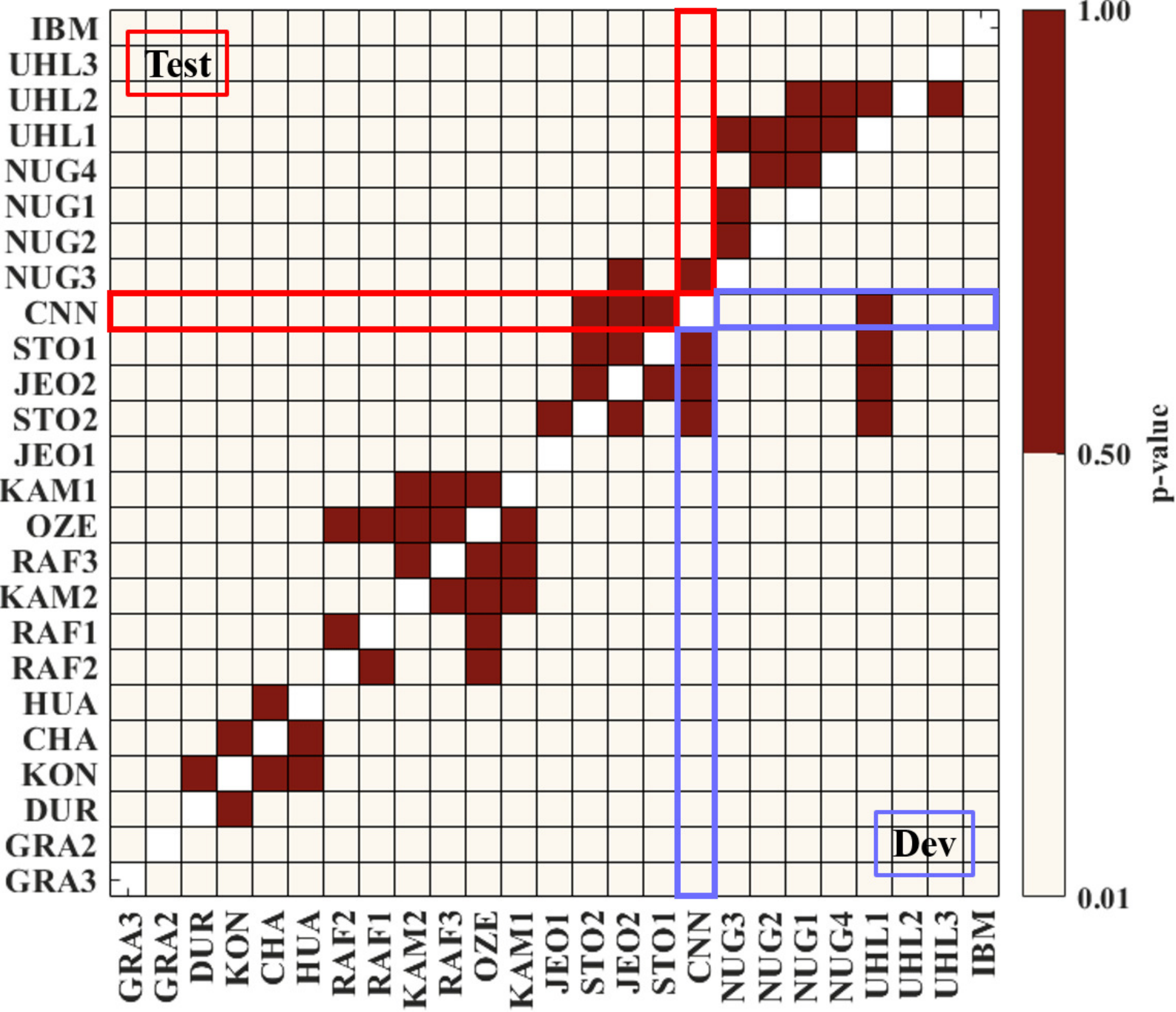}}
	\caption{$P$-values of the Pair-wise difference of Wilcoxon signed-rank test over different pairs of SVS systems. The upper triangle represents the result of the test set and the lower triangle represents the result of the dev set. Values $p > 0.05$ indicate no significant differences between two SVS systems. Note that the Labels of SVS systems are different in these two sub-figures. They are based on the ranking shown in Figure~\ref{fig:DSDResult}.}
    \label{fig:SignRankTest}
\end{figure}

To evaluate the significance of the difference in performance, a pairwise two-tailed Wilcoxon signed-rank test with Bonferroni correction~\cite{EUSIPCO16:SignTest} was performed. Figure~\ref{fig:SignRankTest} summarizes the results. There is no statistical difference, in terms of separation quality of the singing voice, between our CNN, UHL(1,2), and NUG(1-4). This relativizes the importance of Figure~\ref{fig:DSDResult} . The only significant different is with UHL3, which uses model blending between UHL1 and UHL2. This results suggests that our CNN might be a suitable candidate for blending with other state-of-the-art systems.

\citet{ISMIR2017:UNet} reported a remarkable performance by using their U-Net architecture trained on a huge industry dataset. We refrained from directly comparing our CNN with the U-net as we are not able to replicate their extraordinary performance when training on the smaller iKala and DSD100 training set. Nevertheless, by looking the empirical results\footnote{For iKala, the GNSDRs for both singing voice and music accompaniment are $9.50$ dB and $6.34$ dB respectively; For DSD100, the SDRs for both singing voice and music accompaniment are $2.83$ dB and $6.71$ dB respectively.} reported by similar U-nets~\cite{stoller2017adversarial,stoller2018jointly}, we are confident that our CNN is able to compete with the U-net architecture.

\section{Conclusion}
\label{sec:conclusion}

A singing voice separation model inspired by recent advances in image processing, e.g. pixel-wise image classification, is presented in this paper. Details of the full design process of this model are given, including preprocessing steps such as how the mixture signal can be transformed to form the model's input. The full architecture of the proposed convolutional neural network is discussed, which includes an Ideal Binary Mask component as the prediction target label. Our unique network approach includes IBM target labels, cross entropy loss, and pretraining the CNN as an autoencoder on singing voice spectrogram segments. 

Computational results based on the iKala and DSD100 dataset show that the proposed system can compete with cutting-edge voice separation systems. On the iKala dataset, our model reaches $2.2702\mkern-2mu\sim\mkern-2mu 5.9563$ dB Global GNSDR gain over the two best performing algorithms~\cite{LVAICA2017:MIREX20161stRunnerUp,MIREX14:Winner}. Second, on the DSD100 dataset, no statistically significant difference was found between the proposed model and current state-of-the-art (non-fused) systems~\cite{SiSEC2016:DSD100}. Audio examples resulting from this paper are available online\footnote{\url{https://kinwahedwardlin.wordpress.com/}}, together with the spectrogram plots, source code and trained models. 
 
In future research, it would be interesting to further improve the quality of the separated music accompaniment, e.g., by dedicated training on specific instruments in the music accompaniment, and systematically studying the effect of the model's components on the separation quality, such as the choices for the number of feature maps in each layers.

\section{Conflict of Interest Statement}

The authors of this manuscript certify that they have NO affiliations with or involvement in any organization or entity with any financial interest (such as honoraria; educational grants; participation in speakers’ bureaus; membership, employment, consultancies, stock ownership, or other equity interest; and expert testimony or patent-licensing arrangements), or non-financial interest (such as personal or professional relationships, affiliations, knowledge or beliefs) in the subject matter or materials discussed in this manuscript.

% BibTeX users please use one of
%\bibliographystyle{spbasic}      % basic style, author-year citations
%\bibliographystyle{spmpsci}      % mathematics and physical sciences
%\bibliographystyle{spphys}       % APS-like style for physics
%\bibliographystyle{splncs}
\bibliographystyle{splncsnat}  %changed to be able to use citet
\bibliography{ref}   % name your BibTeX data base

\end{document}